\UseRawInputEncoding
\documentclass[a4paper]{article}
\usepackage{amsfonts}
\usepackage{graphicx}
\usepackage{amssymb}
\usepackage{amsmath}
\usepackage{mathrsfs}
\usepackage{latexsym}
\usepackage{geometry}
\usepackage{authblk}
\usepackage{verbatim}
\usepackage{subfigure}
\usepackage{hyperref}
\usepackage{epstopdf}
\usepackage{color}
\usepackage{cite}

\newcommand{\red}[1]{{\color{red} #1}}
\newcommand{\bean}{\begin{eqnarray}}
\newcommand{\eean}{\end{eqnarray}}
\newcommand{\eqs}[1]{Eqs. (\ref{#1})}
\newcommand{\eq}[1]{Eq. (\ref{#1})}
\newcommand{\meq}[1]{(\ref{#1})}

\newcommand{\tab}[1]{Tab. \ref{table1}}
\newcommand{\ppa}[2]{\left(\frac{\partial}{\partial #1}\right)^{#2}}

\newcommand{\ppn}[2]{\frac{\partial #1}{\partial #2}}

\newcommand{\bea}{\begin{eqnarray*}}
\newcommand{\eea}{\end{eqnarray*}}

\newcommand{\eqn}{&=&}
\newcommand{\non}{\nonumber \\}

\begin{document}
\title{General mass formulas for charged Kerr-AdS black holes}
\author{Yunjiao Gao\thanks{yunjiao@mail.bnu.edu.cn}}
\author{Zhenbo Di\thanks{dizhenbo@mail.bnu.edu.cn}}
\author{Sijie Gao\thanks{Corresponding author: sijie@bnu.edu.cn}}
\affil[1]{Department of Physics, Beijing Normal University, Beijing , China,100875}

\maketitle
\begin{abstract}
It is well-known that the mass of a non-asymptotically flat spacetime cannot be  uniquely defined. Some mass formulas for the Kerr-AdS black hole have been found and used in studying black hole thermodynamics. However, the derivations usually need a background subtraction to eliminate the divergence at infinity. It is also unknown  whether the mass depends on the choice of coordinates. In this paper, we provide a more straightforward derivation for the mass formula, only demanding that the first law of black hole thermodynamics and Smarr formula are satisfied. We first make use of the Iyer-Wald formalism to derive a first law which avoids the divergence at infinity. Then we apply this formula to charged Kerr-AdS black hole expressed in the coordinates rotating at infinity.
However, the first law associated with the timelike Killing vector field $\frac{\partial}{\partial t}$  is  not integrable. Then, by making use of the gauge freedom of $t$,  we find a favorite parameter $t'$ which just makes the mass integrable. Applying the scaling argument, we show that the mass satisfies the Smarr formula and takes the form $M/\Xi^{3/2}$. Moreover, applying the conformal method with $\ppn{}{t'}$,  we obtain the same mass. By applying the first law to the coordinates which is not rotating at infinity, we find a preferred time $T$ that makes the first law integrable and the mass is just the  familiar mass $M/\Xi^2$ in the literature. This mass is also confirmed by the  conformal method. We find that the two mass formulas correspond to different families of observers and the preferred Killing times. So our work clarifies the ambiguities of mass in Kerr-AdS spacetimes.

\end{abstract}

\section{Introduction}\label{sec1}
The definition of mass in a non-asymptotically spacetime is always a problem since the ADM mass fails and the Komar mass becomes divergent \cite{Deruelle:2005qk} in general at infinity. As a typical non-asymptotically flat spacetime, the mass of Kerr-dS(AdS) spacetime has been studied and different solutions have been proposed.
Magnon  eliminated the divergent term in the Komar mass  by subtracting the suitably chosen reference background \cite{magnon1985komar}. By introducing an anti-symmetric Killing potential $\omega^{a b}$,
Kastor obtained a Komar-like mass formula \cite{Kastor:2009wy,Kastor:2008xb}, where the divergent term has been canceled by the term containing $\omega$. Awad and Johnson defined the conserved charge by using the Euclidean action\cite{Awad:1999xx}. Since the original Euclidean action is divergent,  suitable counterterms have to be added to regularize divergent integrals. Ashtekar \cite{Ashtekar:1984zz} and Das \cite{ Das:2000cu}  made use of the conformally rescaled metric $\bar{g}_{\mu \nu}=\Omega^2 g_{\mu \nu}$, and its Weyl tensor ${{\bar{C}}^{\mu}} _{\nu \rho \sigma}$ to define the conserved charge, where $\Omega$ is the conformal factor. It is worth remarking that the conformal method directly gives a finite result, unlike the previous methods required to eliminate divergence. The above methods give the same mass expression $\frac{M}{\Xi}$ for the Kerr-AdS black hole expressed in coordinates comoving with rotating observers at infinity , where $M$ is the mass parameter in the metric and $\Xi=1-a^2/l^2$ with $l$ being the AdS radius and $a$ being the angular momentum per unit mass. However, it is easy to check that this mass formula does not satisfy the first law of thermodynamics in the normal (without considering the variation of the cosmological constant  $\Lambda$) or extended phase space (considering the variation of $\Lambda$) \cite{Gibbons:2004ai}. Therefore, a more sutiable mass formula was needed.

 Using dimensionless $SO(3, 2)$ generators, Henneaux and Teitelboim \cite{Henneaux:1985tv} first derived the mass $\tilde M=\frac{M}{\Xi^2}$, which  satisfies the first law of black hole thermodynamics. In Ref.\cite{Caldarelli:1999xj}, the authors  also obtained the mass $\frac{M}{\Xi^2}$ by using the  above Komar mass and Euclidean action method. However, they used the timelike Killing vector field $\frac{1}{\Xi}\frac{\partial}{\partial t}$ to calculate the Komar mass and conserved charges and used $\frac{\partial}{\partial t}+\Omega_H\frac{\partial}{\partial\phi}$ to define the surface gravity. This does not seem to be consistent. Using conserved Killing charges \cite{Deser:2002jk}, Deser et al also obtained the mass $\frac{M}{\Xi^2}$ \cite{Deser:2005jf}.
  The result $\frac{M}{\Xi^2}$ has been widely used in the literature \cite{Altamirano:2014tva,Zeng:2019aao,Tang:2020zhq,Wu:2023sue,Li:2020nnz,Gunasekaran:2012dq,Dolan:2012jh,Okcu:2017qgo} to study  thermodynamics in the extended phase space.
   Recently, Hajian and M. M. Sheikh-Jabbari \cite{Hajian:2015xlp,Hajian:2016kxx} derived the first law by applying the Iyer-Wald formalism and then obtained  the mass $\frac{M}{\Xi^2}$ by integration. Note that this first law does not include the variation of $\Lambda$, thus not applicable to the extended phase space \cite{Kubiznak:2016qmn,Pradhan:2016rff,Guo:2021ere,Kruglov:2022lnc,Lobo:2021bag,Liang:2020uul,Yao:2020ftk}.

Concerning the ambiguities of mass in the Kerr-AdS spacetime, there are two major issues that need to be clarified. Firstly, the Kerr-AdS metric can be expressed in two coordinate systems, corresponding to rotating and non-rotating observers at infinty. Is the mass different for those coordinates (observers)?
Secondly, for each family of coordinates, the mass depends on the choice of the timelike Killing vector field $\ppn{}{t}$. Since the spacetime is not asymptotically flat, one can obtain another Killing vector field by multiplying $\ppn{}{t}$ with any constant. Can we fix the gauge freedom?
In this paper, we require that the mass should  satisfy both the first law in the extended space and the Smarr formula. By this natural criteria, we shall see that the above questions can be clearly answered.
We first apply the Iyer-Wald formalism to obtain the first law of charged Kerr-AdS black holes in the extended  space. However, we find that the mass associated with the customary $\frac{\partial}{\partial t}$ in this first law is not integrable for the coordinates rotating at infinity. As we mentioned above, there is a gauge freedom for Killing time $t$. Taking this into account, we modify the first law and find that there does exist a special Killing time $t'$, which makes mass in the first law integrable.  Then by applying the scaling transformation \cite{Sudarsky:1993kh} proposed by Wald to the modified first law, we obtain the mass as $\hat{M}=\frac{M}{\Xi^{3/2}}$ which satisfies the Smarr formula. As an independent check, we apply the conformal technique to the metric with the same Killing time $t'$  and recover the mass $\frac{M}{\Xi^{3/2}}$. This consistency provides a strong support for our result.
We then consider the metric described by the coordinates associated with nonrotating observers at infinity. We find there is a preferred  timelike Killing vector field $\ppn{}{T}$ that leads to the familiar mass formula $\frac{M}{\Xi^2}$. This mass also satisfies the first law and Smarr formula. Again, this result is confirmed by the conformal method.

The rest of this paper is organized as follows. In section \ref{sec2}, we derive the  first law  in the extended phase space by using the Iyer-Wald formulism. In section \ref{sec3}, by applying the first law to the charged Kerr-AdS metric expressed in coordinates rotating at infinity and choosing a preferred Killing time, we derive a mass formula $M/\Xi^{3/2}$, which satisfies both the first law and Smarr formula. We also check that this formula is consistent with what is obtained from the conformal method.  In section \ref{sec4}, we apply the same first law to the metric expressed in coordinates non-rotating at infinity and derive the mass formula $M/\Xi^2$. Again, the conformal method gives the same result. Conclusions are given in section \ref{sec6}. We adopt the geometric units $c=G=1$ through this paper.

\section{The first law of thermodynamics in the extended phase space}\label{sec2}
Our purpose is to find the first law for a black hole with cosmological constant $\Lambda$, which is associated with the Lagrangian
\cite{Caldarelli:1999xj}
  \begin{equation}
\mathbf{L}=\frac{1}{16 \pi}\left(R-2\Lambda-F^{ab}F_{ab}\right) \boldsymbol{\epsilon},\label{cre}
\end{equation}
where $R$ is the Ricci scalar, $\boldsymbol{F} =d\boldsymbol{A}$ is the electromagnetic field tensor of the vector potential $\boldsymbol{A}$ ($d$ denotes the\emph{ exterior differentiation operator}), and $\boldsymbol{\epsilon}$ is the four-dimensional volume element. Here and below, we  use bold letters to denote differential forms, i.e., antisymmetric tensors.
 The variation of $\mathbf{L}$ leads to \footnote{The derivations of \eqs{bxpo}-\meq{yw} are demonstrated in Appendix \ref{appb}.} \cite{Iyer:1994ys,Iyer:1995kg}
  \begin{equation}
\delta \mathbf{L}=\mathbf{E_g} \delta g^{ab}+\mathbf{E_{A}}\delta A_f+d \boldsymbol{\Theta} \left(\phi, \delta \phi\right)-\frac{1}{8 \pi}\boldsymbol{\epsilon}\delta\Lambda,\label{bxpo}
\end{equation}
where
\bean
\mathbf{E_g}&=&\frac{1}{16 \pi}\boldsymbol{\epsilon} \left(G_{ab}-8\pi T_{ab}^{\rm{EM}}+ \Lambda g_{ab}\right), \label{mtr}\\
\mathbf{E_{A}}&=&\frac{1}{4 \pi}\nabla_e F^{ef}\boldsymbol{\epsilon},\label{mtr2}
\eean
with
\begin{equation}
T_{ab}^{\rm{EM}}=\frac{1}{4\pi}\left(F_{ac}{F_{b}}^{c}-\frac{1}{4}g_{ab}F_{cd}F^{cd}\right)
\label{klh}
\end{equation}
being the electromagnetic stress-energy tensor.

$\mathbf{E_g}=\mathbf{E_{A}}=0$ yields the equations of motion. Note that $\Lambda$ is usually a constant in the Lagrangian. Variation of $\Lambda$ has been widely used in the context of extended phase space. In Ref.\cite{cqg09}, $\Lambda$ is called ``working variable'' and its variation does not yield the equation of motion, unlike the usual ``dynamical variable''.
The boundary term
\begin{equation}
\boldsymbol{\Theta}\left(\phi, \delta \phi\right)= \frac{1}{16 \pi} \epsilon_{d a b c} g^{d e} g^{f h}\left(\nabla_f \delta g_{e h}-\nabla_e \delta g_{f h}\right)-\frac{1}{4 \pi}\epsilon_{e a b c}F^{ef}\delta A_f \label{yw}
\end{equation}
is called the symplectic potential 3-form \cite{Iyer:1994ys}.
Define the Noether current\cite{Wald:1993nt,Iyer:1994ys}
\bean
\mathbf{J}=\boldsymbol{\Theta}\left(\phi, \mathcal{L}_{\xi} \phi\right)-\xi \cdot \mathbf{L},\label{yuee}
\eean
where $\xi^a$ is a Killing vector field. $\xi \cdot \mathbf{L}$ denotes the contraction of  $\xi^a$ with the first index of $\mathbf{L}$.
In addition, $\mathbf{J}$ can be written as\cite{Iyer:1995kg}
\bean
\mathbf{J}=\mathbf{C}_\xi+d \mathbf{Q}_\xi,\label{zbyv}
\eean
where the  $2$-form $\mathbf{Q}$ is referred to as the Noether charge.

Substituting \eqs{cre} and \meq{yw} into \eqs{yuee} and \meq{zbyv},
we have
\begin{align}
\mathbf{Q}_\xi&=-\frac{1}{16 \pi} \epsilon_{a b c d} \nabla^c \xi^d-\frac{1}{8 \pi}\xi^h A_h F^{cd}\epsilon_{cdab},\label{beo}\\
\mathbf{C}_{\xi}
&=\frac{1}{8 \pi}\xi^e \left({G^d}_e +\Lambda {g^d}_e+8\pi\left(A_e j^d-{{T^{\rm{EM}}}_e}^d\right) \right)\epsilon_{dabc}\label{red},
\end{align}
where $\nabla^a F_{ba}=4\pi j_b$ has been used.
\eqs{mtr}-\meq{mtr2} and \meq{red} indicate that when $\mathbf{E_g}=\mathbf{E_{A}}=0$, i.e., the equations of motion are satisfied,  $\mathbf{C}_{\xi}$ vanishes. Hence \eq{zbyv} becomes
\begin{equation}
\mathbf{J}=d \mathbf{Q}_\xi. \label{vser}
\end{equation}
Defining the variation of the Noether current $\meq{yuee}$ as
 \bean
\bar{\delta}\mathbf{J}&=\delta\boldsymbol{\Theta}\left(\phi, \mathcal{L}_{\xi} \phi\right)-\xi \cdot \delta \mathbf{L}\label{vop}
\eean
where  $\bar{\delta}$ denotes variation with $\xi^a$ being fixed \cite{Gao:2003ys}, as opposed to the complete variation $\delta$.
Making use of \eq{bxpo}, $\mathbf{E_g}=\mathbf{E_{A}}=0$, and $\delta\boldsymbol{\Theta}\left(\phi, \mathcal{L}_{\xi} \phi\right)=\mathcal{L}_{\xi} \boldsymbol{\Theta}\left(\phi, \delta \phi\right)=\xi \cdot d \boldsymbol{\Theta}\left(\phi, \delta \phi\right)+d(\xi \cdot \boldsymbol{\Theta}\left(\phi, \delta \phi\right))$, \eq{vop} is rewritten as
\bean
\bar{\delta}\mathbf{J}=d (\xi \cdot\boldsymbol{\Theta}\left(\phi, \delta \phi\right))+\frac{1}{8 \pi}\xi \cdot\boldsymbol{\epsilon}\delta\Lambda. \label{rest}
\eean
Substituting \eq{vser} into \eq{rest}, we have
\bean
d\left(\bar{\delta}\mathbf{Q}_{\xi}-\xi \cdot\boldsymbol{\Theta}\left(\phi, \delta \phi\right)\right)=\frac{1}{8 \pi}\xi \cdot\boldsymbol{\epsilon}\delta\Lambda.\label{esf}
\eean
\eq{esf} has been obtained in Ref.\cite{Couch:2016exn}.

Now we make use of  \eq{esf} to derive the first law in the extended space. Choosing a spacelike hypersurface $\Sigma$ connecting the horizon $H$ and any two-dimensional closed cross-section  $S(t, r=constant$, e.g. $t=t_0, r=r_0$) surrounding the black hole and applying Stokes' theorem to \eq{esf}, we have
\bean
\int_{S}\left(\bar{\delta}\mathbf{Q}_{\xi}-\xi \cdot\boldsymbol{\Theta}\left(\phi, \delta \phi\right)\right)-\int_{H}\left(\bar{\delta}\mathbf{Q}_{\xi}-\xi \cdot\boldsymbol{\Theta}\left(\phi, \delta \phi\right)\right)=\int_{r=0}^{r=r_0}\frac{1}{8 \pi}\xi \cdot\boldsymbol{\epsilon}\delta\Lambda-\int_{r=0}^{r=r_h}\frac{1}{8 \pi}\xi \cdot\boldsymbol{\epsilon}\delta\Lambda,\label{cny}
\eean
where $r_h$ is the horizon radius.  
  Following \eq{cny}, we set $\xi^a=t^a=\left(\frac{\partial}{\partial t}\right)^a$, and define the variation of the mass  as
\bean
\delta \mathcal{M}=\int_{S}\left(\bar{\delta}\mathbf{Q}_{t}-t \cdot\boldsymbol{\Theta}\left(\phi, \delta \phi\right)\right)-\int_{r=0}^{r=r_0}\frac{1}{8 \pi}t \cdot\boldsymbol{\epsilon}\delta\Lambda.\label{xm}
\eean
It follows from \eq{cny} that this definition is  independent of the choice of $S$. Thus, the divergence of integral at infinity in previous treatments can be avoided.
Similarly, by taking $\xi^a=\phi^a=\left(\frac{\partial}{\partial \phi}\right)^a$, we define the variation of the angular momentum as
\bean
\delta J\eqn-\int_{S}\left(\bar{\delta}\mathbf{Q}_{\phi}-\phi \cdot\boldsymbol{\Theta}\left(\phi, \delta \phi\right)\right)+\int_{r=0}^{r=r_0}\frac{1}{8 \pi}\phi \cdot\boldsymbol{\epsilon}\delta\Lambda \non
\eqn -\int_{S}\bar{\delta}\mathbf{Q}_{\phi},\label{angularr}
\eean
where we have used the fact that the pullbacks of $\phi\cdot \boldsymbol{\Theta}$ and  $\phi\cdot \boldsymbol{\epsilon}$ vanish.
In the following, we shall adopt the gauge \cite{Bardeen:1973gs} $\delta t^a=0, \delta \phi^a=0$. Therefore, $\bar{\delta}\mathbf{Q}_{\phi}=\delta\mathbf{Q}_{\phi}$ and
 $J$ takes the form
\begin{equation}
 J=-\int_{S}\mathbf{Q}_{\phi}=\frac{1}{16 \pi}\int_{S} \epsilon_{a b c d} \nabla^c \phi^d+\frac{1}{8 \pi}\int_{S}\phi^h A_h F^{cd}\epsilon_{cdab}.\label{utyr}
\end{equation}
It follows \eq{cny} that $J$  is also  independent of the choice of $S$.

Define the Killing vector field $\xi^a=t^a+\Omega_{H}\phi^a$ which is normal to the horizon, where $\Omega_{H}=-\frac{g_{t\phi}}{g_{\phi\phi}}|_H$ is the horizon angular velocity. Then, we have
\bean
\delta \xi^a=\phi^a\delta\Omega_{H} \label{zlv}.
\eean
Combining \eqs{cny}, \meq{xm} and \meq{angularr}, we have
\begin{equation}
\delta \mathcal{M}-\Omega_H \delta J=\int_{H}\left(\bar{\delta}\mathbf{Q}_{\xi}-\xi \cdot\boldsymbol{\Theta}\left(\phi, \delta \phi\right)\right)-\int_{r=0}^{r=r_h}\frac{1}{8 \pi}t \cdot\boldsymbol{\epsilon}\delta\Lambda\label{vt},
\end{equation}
where we have taken advantage of the fact that $\phi\cdot \boldsymbol{\epsilon}$ vanishes on $H$. Next, we calculate the right hand side of \eq{vt}. For the first term, considering the difference between variations $\delta$ and $\bar\delta$, we have \cite{Gao:2003ys}
\bean
\int_{H}\bar{\delta}\mathbf{Q}_{\xi}=\int_{H}\delta\mathbf{Q}_{\xi}-\int_{H}\mathbf{Q}_{\delta\xi}.\label{wb}
\eean
Substituting \eq{beo} into \eq{wb}, and using  \eqs{utyr}-\meq{zlv}, the relation\cite{Wald:1984rg},
\begin{equation}
\int_{H}\epsilon_{abcd}\nabla^{c}\xi^{d}= -2\kappa A,
\end{equation}
and definitions of the electric potential and charge
\begin{align}
\Phi_{H}&\equiv-A_a \xi^a|_H,\label{loo}\\
Q&\equiv\frac{1}{8 \pi}\int_{H} F^{cd}\epsilon_{abcd},\label{nlk}
\end{align}
 we have
\bean
\int_{H}\bar{\delta}\mathbf{Q}_{\xi}=\delta\left(\frac{\kappa A}{8 \pi}+\Phi_h Q\right)+J\delta\Omega_{H},\label{xmu}
\eean
where  $\kappa$ is the surface gravity defined by $\kappa=n^a\xi^b\nabla_{a}\xi_{b}$ with $n_a \xi^a=-1$.
The variation of $\kappa$ is
\cite{Bardeen:1973gs}
\begin{equation}
\delta\kappa=-\frac{1}{2}\xi^b\nabla^a\delta g_{ab}+n^a\xi_b\nabla_a\phi^b\delta\Omega_{H}.\label{pht}
\end{equation}
For the second term in the right hand side of \eq{vt}, using \eqs{yw} and \meq{pht},$\epsilon_{a b c d}=\epsilon_{a b} \wedge \xi_{c} \wedge n_{d}$ on $H$  and the fact that $\xi_{e} F^{ef}\propto \xi^f$ for the stationary black hole \cite{Gao:2001ut,Gao:2003ys}, we obtain
\bean
\int_{H}\xi \cdot\boldsymbol{\Theta}\left(\phi, \delta \phi\right)=\frac{1}{8 \pi}A\delta\kappa+\frac{1}{16 \pi}\int_{H}\epsilon_{bcfd}\nabla^f\phi^d\delta\Omega_{H}+\frac{1}{4 \pi}\int_{H}\epsilon_{bc} F^{eh}\xi_{h}n_e\xi^f\delta A_f.\label{pl}
\eean
Multiplying the variation of \eq{loo} by \eq{nlk}, we have
\bean
Q\delta\Phi_{H}
=-\frac{1}{8 \pi}\int_{H} F^{cd}\epsilon_{abcd}A_e \phi^e\delta\Omega_{H}+\frac{1}{4 \pi}\int_{H} F^{dc}\epsilon_{a b} \xi_{c} n_{d}\xi^e\delta A_e\label{bpl},
\eean
where \eq{zlv} and $\epsilon_{a b c d}=\epsilon_{a b} \wedge \xi_{c} \wedge n_{d}$ have been used.
Subtracting  \eq{bpl} by \eq{pl} and using \eq{utyr}, we get
\bean
\begin{aligned}
\int_{H}\xi \cdot\boldsymbol{\Theta}\left(\phi, \delta \phi\right)-Q\delta\Phi_{H}=\frac{1}{8 \pi}A\delta\kappa+J\delta\Omega_{H}.\label{xjl}
\end{aligned}
\eean
Combining \eqs{xmu} and \meq{xjl}, we obtain \cite{Gao:2003ys}
\bean
\int_{H}\bar{\delta}\mathbf{Q}_{\xi}-\xi \cdot\boldsymbol{\Theta}\left(\phi, \delta \phi\right)=\frac{\kappa}{8 \pi} \delta A+\Phi_{H} \delta Q.\label{qp}
\eean
 Making use of \eq{qp} and defining the geometry volume \cite{Cvetic:2010jb}
 \begin{equation}
   \mathcal{V}= \int_{r=0}^{r=r_h}t \cdot\boldsymbol{\epsilon}
   =\int_0^{2\pi}\int_0^\pi  \int_{0}^{r_h}\sqrt{-g}dr d\theta  d\phi \label{mn},
\end{equation}
\bean
P=-\frac{\Lambda}{8 \pi},\label{pwy}
\eean
\eq{vt} can be rewritten as
\begin{equation}
\delta \mathcal{M}=\frac{\kappa}{8 \pi} \delta A+ \mathcal{V}\delta P+\Omega_H \delta J+\Phi_{H} \delta Q\label{vnpe},
\end{equation}
which is the first law with cosmological constant in the extended phase space. Compared with previous works \cite{Kastor:2009wy,Dolan:2013ft,Hyun:2017nkb}, we derive the first law without introducing an anti-symmetric Killing potential $\omega^{a b}$ to cancel the divergent term.
\section{Mass of the Charged Kerr-AdS black hole associated with coordinates rotating at infinity }\label{sec3}
This section is divided into two subsections. In subsection \ref{sec3.1},  we shall apply the first law \meq{vnpe} to charged Kerr-AdS black holes expressed in coordinates which  is rotating at infinity and obtain the mass of the spacetime. In subsection \ref{sec3.2},  we calculate the conformal mass and find that it is the same as the thermodynamic mass.

\subsection{Deriving the mass from the first law} \label{sec3.1}
Consider a charged Kerr-AdS black hole described by coordinates which are rotating at infinity. The metric is given by \cite{Caldarelli:1999xj}
\begin{equation}
d s^{2}=-\frac{\Delta_{r}}{\rho^{2}}\left[d t-\frac{a \sin ^{2} \theta}{\Xi} d \phi\right]^{2}+\frac{\rho^{2}}{\Delta_{r}} d r^{2}+\frac{\rho^{2}}{\Delta_{\theta}} d \theta^{2}+\frac{\Delta_{\theta} \sin ^{2} \theta}{\rho^{2}}\left[a d t-\frac{r^{2}+a^{2}}{\Xi} d \phi\right]^{2},\label{ur}
\end{equation}
where
\begin{equation}
\rho^{2}=r^{2}+a^{2} \cos ^{2} \theta, \quad \Xi=1+\frac{\Lambda a^{2}}{3},\quad \Lambda=\frac{-3}{l^2}\label{ruf},
\end{equation}
\begin{equation}
\Delta_{r}=\left(r^{2}+a^{2}\right)\left(1-\frac{\Lambda r^{2}}{3}\right)-2 M r+q^2, \quad \Delta_{\theta}=1+\frac{\Lambda a^{2}}{3} \cos ^{2} \theta.
\end{equation}
The corresponding electromagnetic vector potential takes the form
\begin{equation}
A_a=-\frac{q r}{\rho^2 }\left(\mathrm{d} t_a-\frac{a \sin ^2 \theta}{\Xi} \mathrm{d} \phi_a\right)\label{vf}.
\end{equation}
The surface gravity $\kappa$,  the area $A$ of the event horizon, and the angular velocity $\Omega_H$  of the black hole are given by\cite{Caldarelli:1999xj}
\begin{align}
\kappa=&\frac{\left(3 r_h^4 + a^2 \left(-l^2 + r_h^2\right) + l^2 \left(-q^2 + r_h^2\right)\right)}{2 l^2 r_h \left(a^2 + r_h^2\right)}\label{vyh},\\
A=&\frac{4\pi\left(a^2 + r_h^2\right)}{\Xi},\quad\Omega_H=\frac{a \Xi}{\left(a^2 + r_h^2\right)}\label{nup}.
\end{align}
The other quantities related to the first law can be obtained by substituting the metric \meq{ur} and vector potential \meq{vf} into \eqs{utyr}, \meq{loo}, \meq{nlk}, \meq{mn} and \meq{pwy}. Then, we have\cite{Caldarelli:1999xj}
\bean
J\eqn\frac{M a}{\Xi^2},\label{ge}\\
\Phi_H\eqn\frac{r_h q }{a^2 + r_h^2},\label{bh}\\
Q\eqn\frac{q}{\Xi} ,\label{kox} \\
\mathcal{V}\eqn\frac{4\pi l^2  r_h \left(a^2 + r_h^2\right)}{3 \left(-a^2 + l^2\right)}, \label{ckd}\\
P\eqn\frac{3}{8\pi l^2}\label{xbup}.
\eean

Substituting \eqs{vyh}-\meq{xbup} into \eq{vnpe}, we have
\bean
\begin{aligned}
\delta \mathcal{M}=&\left[\frac{\left(3 r_h^4 + a^2 \left(-l^2 + r_h^2\right) + l^2 \left(-q^2 + r_h^2\right)\right)}{16\pi l^2 r_h \left(a^2 + r_h^2\right)} \delta \left(\frac{4\pi(a^2 + r_h^2)}{\Xi}\right)+\right.\\&\left. \frac{4\pi l^2  r_h \left(a^2 + r_h^2\right)}{3 \left(-a^2 + l^2\right)}\delta \left(\frac{3}{8\pi l^2}\right)+\frac{a \Xi}{\left(a^2 + r_h^2\right)} \delta \left(\frac{M a}{\Xi^2}\right)+\frac{r_h q }{a^2 + r_h^2}\delta\frac{q}{\Xi}\right].\label{xpz}
\end{aligned}
\eean
The horizon of the charged Kerr-AdS black hole \meq{ur} is decided by $\Delta_r=0$, which gives
\bean
 M=\frac{r_h^4+a^2 l^2 + l^2 q^2 + a^2 r_h^2 + l^2 r_h^2 }{2 l^2 r_h}.\label{lzg}
\eean
 The variation of \eq{lzg} leads to
 \bean
\delta r_h=-\frac{-2 a l r_h \left(l^2 + r_h^2\right) \delta a + 2 r_h \left(a^2 r_h^2 + r_h^4\right) \delta l +
 2 l^3 r_h^2 \delta M - 2 l^3 q r_h \delta q}{l \left(-3 r_h^4 + a^2 \left(l^2 - r_h^2\right) + l^2 \left(q^2 - r_h^2\right)\right)}.\label{hpsr}
 \eean
Finally, substituting \eqs{ruf} and \meq{hpsr} into \eq{xpz}, we can express $\delta \mathcal{M}$ as a linear combination of $\delta M$, $\delta a$ and $\delta l$:
 \bean
 \begin{aligned}
 \delta \mathcal{M}=&\frac{l^2}{-a^2 + l^2}\delta M+\frac{3 a \left(r_h^4 + a^2 \left(l^2 + r_h^2\right) + l^2 \left(q^2 + r_h^2\right)\right)}{2 \left(a^2 - l^2\right)^2 r_h}\delta a\\&-\frac{3 a^2 \left(r_h^4 + a^2 \left(l^2 + r_h^2\right) + l^2 \left(q^2 + r_h^2\right)\right)}{2 l \left(a^2 - l^2\right)^2 r_h}\delta l.\label{txgr}
 \end{aligned}
 \eean
 Making using of \eq{lzg}, \eq{txgr} can be simplified as
\bean
\delta \mathcal{M}=C_{M}\delta M+C_{a}\delta a+C_{l}\delta l,\label{vpe}
\eean
where
\bean
 C_{M}=\frac{l^2}{-a^2 + l^2},\quad
 C_{a}=\frac{3 a l^2 M}{\left(a^2 - l^2\right)^2},\quad
 C_{l}=-\frac{3 a^2 l M}{\left(a^2 - l^2\right)^2}. \label{pdw}
\eean
\eq{vpe} is not integrable since $\frac{\partial C_{M}}{\partial a}\neq\frac{\partial C_{a}}{\partial M}$. Note that the mass formula is associated with the timelike Killing vector $\ppa{t}{a}$ which is
not  unique for a non-asymptotically flat spacetimes.  Our next task is to seek the best parameterization for the first law.

We first multiply both sides of \eq{vnpe} by a constant $\beta$, i.e.
\bean
\beta\delta \mathcal{M}=\beta\left(\frac{\kappa}{8\pi}\delta A+\Omega_{H}\delta J+ \mathcal{V} \delta P+\Phi_{H} \delta Q\right).\label{eg}
\eean
Defining
\bean
\delta \hat M=\beta\delta \mathcal{M}\label{sc},
\eean
we have
\bean
\delta \hat{M}=\beta\left(\frac{\kappa}{8\pi}\delta A+\Omega_{H}\delta J+\mathcal{V} \delta P+\Phi_{H} \delta Q\right)\label{oy}.
\eean
Substituting \eqs{vpe} and \meq{pdw} into \eq{sc}, we have
\begin{align}
\partial_M \hat{M}&= \frac{l^2 \beta}{-a^2 + l^2},\label{pf}\\
\partial_a \hat{M}&=\frac{3 a l^2 M \beta}{\left(a^2 - l^2\right)^2},\label{ho}\\
\partial_l \hat{M}&=-\frac{3 a^2 l M \beta}{\left(a^2 - l^2\right)^2}\label{kvl}.
\end{align}
Combining \eqs{pf} and \meq{ho}, we can eliminate $\beta$ and obtain
\bean
(a^2 - l^2)\partial_a \hat{M}  + 3 a M \partial_M \hat{M} = 0\label{yuu}.
\eean
To solve this equation, we assume that $\hat M$ takes the form
\bean
\hat{M}=\gamma(a,l)  M \label{eed}.
\eean
A natural requirement is that  $\hat{M}=M$ when $a=0$, which leads to
\bean
\gamma|_{a=0}=1 \label{emd}.
\eean
Then it is not difficult to solve \eq{yuu} and find
\begin{equation} \hat{M}=\frac{M}{\Xi^{3/2}}.\label{ort}
\end{equation}
Substitution of  \eq{ort} into \eq{pf} yields
\bean
\beta=\sqrt{l^2/\left(l^2 - a^2\right)}=\frac{1}{\sqrt{\Xi}}.\label{krw}
\eean
Then one can check that \eq{kvl} is  automatically satisfied.
The above treatment is equivalent to choosing the Killing vector field $\ppa{t'}{a}=\frac{1}{\sqrt{\Xi}}\frac{\partial}{\partial t}$
\footnote{In  \cite{Chrusciel:2015sna}, the authors obtained the same mass $\frac{M}{\Xi^{3/2}}$ by using the Hamiltonian flow generated by the Killing vector field $\frac{1}{\sqrt{\Xi}}\frac{\partial}{\partial t}$. However, the first law and the Smarr formula were not considered, and there is no  variation of $\Lambda$. In Ref \cite{Wu:2022xpp}, the authors found the mass of the four-dimensional Reissner-Nordstrom NUT-charged AdS black hole in the form $\frac{M}{\Xi^{3/2}}$ also by rescaling the  Killing time, where $\Xi$ represents a totally different quantity.},
instead of $\ppa{t}{a}$, to define the mass and thermodynamic quantities. We put $\hat \ $ to  the  quantities associated with $\ppa{t'}{a}$  to distinguish  them with those associated with $\ppa{t}{a}$. Thus, we find
\begin{align}
\hat{\Omega}_H&\equiv -\frac{g_{t'\phi}}{g_{\phi\phi}}|_H=\frac{1}{\sqrt{\Xi}}\Omega_{H},\label{unb}\\
\hat{\xi}^a&=\ppa{t'}{a}+\hat{\Omega}_H\ppa{\phi}{a}, \\
\hat{\kappa}&\equiv\sqrt{-\frac{1}{2}\nabla^a\hat{\xi}^b\nabla_a\hat{\xi}_b}\Big|_H=\frac{1}{\sqrt{\Xi}}\kappa,\label{pxg}\\
\hat{V}&\equiv\int_{r=0}^{r=r_h}\left(\frac{\partial}{\partial t'}\right)^a \epsilon_{abcd}=\frac{1}{\sqrt{\Xi}}\mathcal{V},\label{vv}\\
\hat{\Phi}_H&\equiv-A_a\hat{\xi}^a|_H=\frac{1}{\sqrt{\Xi}}\Phi_{H}\label{cgg}.
\end{align}
Substituting \eqs{krw}-\meq{cgg} into \eq{oy}, we find the first law associated with $\ppa{t'}{a} $ as
\bean
 \delta \hat{M}=\frac{\hat{\kappa}}{8\pi}\delta A+\hat{\Omega}_H\delta J+\hat{V}\delta P+\hat\Phi_H \delta Q.\label{jy}
\eean

Next, we shall apply the scaling transformation \cite{Sudarsky:1993kh} to the first law \meq{jy} to derive the Smarr formula. Assume that the metric transforms as   $g_{a b} \rightarrow \alpha^2 g_{a b}$, where $\alpha$ is a constant. Then
\begin{equation}
\sqrt{-g} \rightarrow \alpha^4 \sqrt{-g}, \quad R \rightarrow \alpha^{-2} R,\quad \Lambda \rightarrow \alpha^{-2} \Lambda,\quad F_{ab}F^{ab}\rightarrow \alpha^{-2}F_{ab}F^{ab}.\label{npxt}
\end{equation}
\eq{npxt} guarantees that the Lagrangian \meq{cre} is invariant up to the overall factor $\alpha^{2}$, which implies the theory is invariant under the scaling transformation \footnote{The transformation of $\Lambda$ is crucial to make the Lagrangian invariant and obtain the Smarr formula below\cite{Awad:2022exw}. We see that without varying $P$ , i.e., $\Lambda$, the first law \meq{jy} still holds, but it cannot yield the Smarr formula.}.
Moreover, the Killing vector field should change as \cite{Zhang:2016ilt}
\begin{equation}
\xi^a \rightarrow \alpha^{-1} \xi^a.
\end{equation}
Hence, the thermodynamic quantities should transform as
\bean
\begin{aligned}
& \hat\kappa \rightarrow \alpha^{-1} \hat\kappa \quad A \rightarrow \alpha^2 A, \quad J \rightarrow \alpha^{2} J ,\quad \hat\Omega \rightarrow \alpha^{-1} \hat\Omega, \quad P \rightarrow \alpha^{-2} P,\quad \hat{V} \rightarrow \alpha^{3} \hat{V},\\&Q\rightarrow\alpha Q,\quad \hat\Phi_H\rightarrow\hat\Phi_H \label{mtre}.
\end{aligned}
\eean

Since the transformations do not change the theory, the new quantities should also satisfy the first law \meq{jy}, i.e.,
\bean
 \delta \left(\alpha^x\hat{M}\right)=\frac{\alpha^{-1}\hat\kappa}{8\pi}\delta \left(\alpha^2 A\right)+\alpha^{-1}\hat\Omega\delta \left(\alpha^2J\right)+\alpha^3\hat{V}\delta \left(\alpha^{-2}P\right)+\hat\Phi_{H} \delta \left(\alpha Q\right).
\eean
Here, we have assumed $\hat M\to \alpha^x \hat M$. Since this equation holds for any $\alpha$, we see immediately that $x=1$ and the coefficient of $\delta\alpha$ gives
\begin{equation}
 \hat{M}=\frac{\hat{\kappa} A}{4 \pi}+2\hat{\Omega}_H J-2P \hat{V}+\hat{\Phi}_H Q,\label{bz}
 \end{equation}
 which is the desired Smarr formula. Through \eqs{vyh}-\meq{xbup}, \meq{lzg}, \meq{ort} and \meq{unb}-\meq{cgg}, it's easy to verify \eq{bz} holds.
Therefore, we obtain the mass \meq{ort} satisfying both  the first law \meq{jy} and the Smarr formula \meq{bz}.

The above conclusion can be directly obtained by performing the
 coordinate transformation
\begin{equation}
t =\frac{1}{\sqrt{\Xi}} t',
\end{equation}
 to metric \meq{ur}. Then the line element becomes
\begin{equation}
d s^{2}=-\frac{\Delta_{r}}{\rho^{2}}\left[\frac{1}{\sqrt{\Xi}}d t'-\frac{a \sin ^{2} \theta}{\Xi} d \phi\right]^{2}+\frac{\rho^{2}}{\Delta_{r}} d r^{2}+\frac{\rho^{2}}{\Delta_{\theta}} d \theta^{2}+\frac{\Delta_{\theta} \sin ^{2} \theta}{\rho^{2}}\left[a \frac{1}{\sqrt{\Xi}}d t'-\frac{r^{2}+a^{2}}{\Xi} d \phi\right]^{2}, \label{lbnm}
\end{equation}
\begin{equation}
A_a=-\frac{q r}{\rho^2 }\left(\frac{1}{\sqrt{\Xi}}\mathrm{d} t'_a-\frac{a \sin ^2 \theta}{\Xi} \mathrm{d} \phi_a\right).\label{vbnl}
\end{equation}
Now $t'$ plays the same role as that $t$ plays in metric \meq{ur}. Thus, we just need to apply \eq{lbnm} to the first law \meq{vnpe} with $\ppa{t}{a}$ being replaced by $\ppa{t'}{a}$. The difference is that the first law associated with $\ppa{t'}{a}$ becomes integrable and leads to the mass formula \meq{ort}.

\subsection{The conformal method}\label{sec3.2}
In this subsection, we shall obtain the mass $M/\Xi^{3/2}$ by the conformal method. Ashtekar \cite{Ashtekar:1984zz,Ashtekar:1999jx} and Das \cite{ Das:2000cu}  have made use of the conformally rescaled metric $\mathring{g}_{ab}=\Omega^2 g_{ab}$, and its Weyl tensor $\mathring{C} _{abcd}$ to define the conformal mass (Please refer to appendix \ref{appa} for the details):
 \begin{equation}
Q[\xi]=-\frac{l}{8 \pi} \oint_{C} \mathring{{\mathcal{E}}} _{ab} \xi^a d \mathring{\Sigma}^b,\label{mk}
\end{equation}
where $\xi^a$ is any infinitesimal asymptotic symmetry (i.e. a conformal Killing
field on boundary), $d \mathring{\Sigma}^b$ is the 3-volume element on the boundary, $C$ is the 2-sphere cross section of the conformal boundary, and
\bean
{\mathring{\mathcal{E}}} _{ab}&=l^2 \Omega^{-1}  {\mathring{C} _{a m b n}} \mathring{n}^m \mathring{n}^n,\label{pzx}
\eean
with
\bean
\mathring{n}_a=\mathring{\nabla}_a \Omega.
\eean

Following Ref. \cite{Ashtekar:1984zz}, we take the conformal factor $\Omega=\frac{1}{r}$ ($\Omega=0$ is the boundary since it corresponds to $r\to\infty$). That is to say, $\mathring{n}_a=-\frac{1}{r^2} (dr)_a$. For convenience, we choose $C$ to be a 2-dimensional surface at infinity with $t=constant$. Then
\bean
d \mathring{\Sigma}^b&=\mathring{u}^b \sqrt{\mathring{h}} d\theta d\phi,\label{nk}
\eean
with $\mathring{u}^b$ being the timelike unit normal to $C$, and $\mathring{h}$ being the determinant of the (hypothetical)induced metric on $C$.
 On $C$, it is straightforward to calculate
 \bean
 \mathring{u}_b=-\sqrt{\frac{-1}{\mathring{g}^{t t}}}(dt)_b, \mathring{u}^b= \mathring{g}^{b a}\mathring{u}_a=-\sqrt{\frac{-1}{\mathring{g}^{t t}}}\left[\mathring{g}^{t t}\left(\frac{\partial}{\partial t}\right)^b+\mathring{g}^{\phi t}\left(\frac{\partial}{\partial\phi}\right)^b\right], \mathring{h}=\mathring{g}_{\theta \theta} \mathring{g}_{\phi\phi}. \label{nch}
\eean
 Choosing the Killing vector $\xi^a=\left(\frac{\partial}{\partial t}\right)^a$ and substituting \eqs{nk} and \meq{nch} into \eq{mk}, we have
  \begin{equation}
Q\left[\frac{\partial}{\partial t}\right]=\frac{l}{8 \pi} \oint_{\infty} \mathring{{\mathcal{E}}} _{ab} \left(\frac{\partial}{\partial t}\right)^a \sqrt{\frac{-1}{\mathring{g}^{t t}}}\left[\mathring{g}^{t t}\left(\frac{\partial}{\partial t}\right)^b+\mathring{g}^{\phi t}\left(\frac{\partial}{\partial\phi}\right)^b\right] \sqrt{\mathring{g}_{\theta \theta} \mathring{g}_{\phi\phi}} d\theta d\phi.\label{phj}
\end{equation}
Through \eq{pzx}, we have
\bean
\mathring{{\mathcal{E}}}_{tt}=\frac{-2 M}{l^2}+O\left[\frac{1}{r}\right],\quad \mathring{{\mathcal{E}}}_{t\phi}=-\frac{2 a M \sin^2
\theta}{a^2-l^2}+O\left[\frac{1}{r}\right]\label{kfd}.
\eean
 Substituting \eq{kfd}, the metric \meq{ur} and $\mathring{g}_{\mu \nu}=(\frac{1}{r})^2 g_{\mu \nu}$ into \eq{phj}, we obtain the conformal mass
 \begin{equation}
Q\left[\frac{\partial}{\partial t}\right]=\frac{M}{\Xi}. \label{lk}
\end{equation}
Note that the mass \meq{lk} associated with $\ppa{t}{a}$  does not satisfy the first law as we have mentioned in the introduction. Now we choose the gauge in \ref{sec3.1}, i.e., $\xi^a=\frac{1}{\sqrt\Xi}\frac{\partial}{\partial t}$. Then \eq{mk} suggests
\bean
Q\left[\frac{1}{\sqrt{\Xi}}\ppn{}{t}\right]=\frac{1}{\sqrt{\Xi}}Q\left[\ppn{}{t}\right]
=\frac{M}{\Xi^{3/2}}\,,
\eean
where \eq{lk} has been used in the last step. We see that the conformal mass just coincides with the thermodynamic mass \meq{ort}. So we can conclude  that the conformal method yields a mass satisfying the first law and Smarr formula only when the correct timelike Killing vector field is chosen.

\section{Mass of the Charged Kerr-AdS black hole associated with coordinates non-rotating at infinity}\label{sec4}
Similarly to section \ref{sec3}, we shall calculate the thermodynamic mass and conformal mass, respectively. The difference is that the coordinates will be associated with non-rotating observers at infinity.

\subsection{Deriving the mass from the first law}\label{sec4.1}
By applying a coordinate transformation to metric \meq{ur} \cite{Gwak:2022xje}
\begin{equation}
t = T, \quad \phi = \Phi+\frac{1}{3} a \Lambda T,\label{ogf}
\end{equation}
we obtain a new observer associated with $\ppa{T}{a}$, which is related to $\ppa{t}{a}$ by
\bean
\ppa{T}{a}=\ppa{t}{a}+\frac{1}{3}a\Lambda\ppa{\phi}{a}\,.
\eean
The new observer is non-rotating  at infinity \cite{Gwak:2022xje} and the metric in these coordinates becomes:
\bean
\begin{aligned}
d s^2= & -\frac{\Delta_r}{\rho^2 \Xi^2}\left(\Delta_\theta d T-a \sin ^2 \theta d \Phi\right)^2+\frac{\rho^2}{\Delta_r} d r^2  +\frac{\rho^2}{\Delta_\theta} d \theta^2 \\
&+\frac{\Delta_\theta \sin ^2 \theta}{\rho^2 \Xi^2}\left\{a\left(1+\frac{r^2}{l^2}\right) d T-\left(r^2+a^2\right) d \Phi\right\}^2\,,\label{kop}
\end{aligned}
\eean
and the corresponding electromagnetic vector potential takes the form
\begin{equation}
A_a=-\frac{q r}{\rho^2 \Xi}\left(\Delta_\theta d T_a-a \sin ^2 \theta d \Phi_a\right) .
\end{equation}
Replacing $\Omega_H$ in \eq{nup} with \cite{Hajian:2016kxx}
\bean
\Omega_H=\frac{a \left(l^2 + r_h^2\right)}{l^2\left(a^2 + r_h^2\right)},
\eean
the physical quantities of the black hole \meq{kop} are still given by \eqs{vyh}-\meq{xbup}.
Then following the strategy of \ref{sec3.1} and substituting these physical quantities into \eq{vnpe}, we can express $\delta \mathcal{M}$ as a linear combination of $\delta M$, $\delta a$ and $\delta l$:
\bean
\delta \mathcal{M}=C_{M}\delta M+C_{a}\delta a+C_{l}\delta l,\label{bne}
\eean
where
\bean
 C_{M}=\frac{l^4}{\left(-a^2 + l^2\right)^2},\quad
 C_{a}=-\frac{4 a l^4 M}{\left(a^2 - l^2\right)^3},\quad
 C_{l}=\frac{a^2 l \left(a^2 + 3 l^2\right) M}{\left(a^2 - l^2\right)^3}. \label{hrs}
\eean
It is not difficult to find that , e.g., $\frac{\partial C_{M}}{\partial a}=\frac{\partial C_{a}}{\partial M}$, $\frac{\partial C_{M}}{\partial l}\neq\frac{\partial C_{l}}{\partial M}$. So \eq{bne} is not integrable. But if we do not consider the variation of $\Lambda$ \cite{Hajian:2016kxx,Chrusciel:2015sna}, i.e. $C_{l}\delta l=0$,  \eq{bne} becomes
\bean
\delta \hat{\mathcal{M}}=\frac{l^4}{\left(-a^2 + l^2\right)^2}\delta M-\frac{4 a l^4 M}{\left(a^2 - l^2\right)^3}\delta a.\label{ufa}
\eean
It is easy to check that $\hat{\mathcal{M}}$ in \eq{ufa} is integrable. Again, we impose that
 $\hat{\mathcal{M}}=M$ at $a=0$ and find
\bean
\hat{\mathcal{M}}=\frac{M}{\Xi^2}.\label{hks}
\eean
The mass \meq{hks} satisfies the first law
\begin{equation}
\delta \hat{\mathcal{M}}=\frac{\kappa}{8 \pi} \delta A+\Omega_H \delta J+\Phi_{H} \delta Q\label{pev}.
\end{equation}
Since $\hat{\mathcal{M}}$ is associated with $\frac{\partial}{\partial T}$, it is natural to ask if the choice of $\ppn{}{T}$ is unique. For this purpose, we use the method of subsection \ref{sec3.1} again and  multiply both sides of \eq{ufa} by a constant $\beta$  ( could depend on $M$ and $a$ )
\bean
\delta\tilde{\mathcal{M}}\equiv\beta\delta \hat{\mathcal{M}}=\beta \left(C_{M}\delta M+C_{a}\delta a\right)=\beta \delta\left(\frac{M}{\Xi^2}\right).\label{pfr}
\eean
We still assume that $\tilde{\mathcal{M}}$  takes the form
\bean
\tilde{\mathcal{M}}=\tilde{\gamma}\left(a,l\right)  M.\label{kps}
\eean
 Substituting \eqs{kps} and \meq{hrs} into \eq{pfr}, we have
 \bean
\tilde{\gamma}\left(a,l\right)\delta M+M \ppn{\tilde{\gamma}\left(a,l\right)}{a} \delta a=\beta \frac{l^4}{\left(-a^2 + l^2\right)^2}\delta M-\beta \frac{4 a l^4 M}{\left(a^2 - l^2\right)^3}\delta a. \label{xzr}
\eean
From the coefficient of $\delta M$, we get immediately
\bean
\ppn{\beta}{M}=0 \,. \label{ppbm}
\eean
If $\tilde{\mathcal{M}}$ is integrable, \eq{pfr} implies
\bean
\frac{\partial{\left(\beta C_{M}\right)}}{\partial a}=\frac{\partial\left(\beta C_{a}\right)}{\partial M}.
\eean
Namely,
\bean
\beta\frac{\partial{ C_{M}}}{\partial a}+C_{M}\frac{\partial{\beta}}{\partial a}=\beta\frac{\partial C_{a}}{\partial M}, \label{bcmb}
\eean
where \eq{ppbm} has been used.
Since $\frac{\partial C_{M}}{\partial a}=\frac{\partial C_{a}}{\partial M}$, it follows from \eq{bcmb} that
\bean
\ppn{\beta}{a}=0  \label{pbaz}
\eean
\eq{ppbm} and \meq{pbaz} suggest that $\beta$ is a trivial factor, independent of black hole parameters. Therefore, $\ppn{}{T}$ is unique (up to a trivial factor) for the first law. In other words, any non-trivial reparametrization of $T$ can not lead to a thermodynamic mass. Thus, associated with non-rotating observers, the mass in \eq{hks} is unique.

Although we have obtained the mass without varying $\Lambda$, it is necessary to vary it in order to obtain the Smarr formula as we have mentioned in footnote 3. To make \eq{bne} integrable, we subtract $V_{int}\delta P$ from both sides of \eq{bne} \cite{Astorino:2016hls,Gao:2022ckf}

\bean
\delta \mathcal{M}-V_{int}\delta P=\frac{\kappa}{8\pi}\delta A+\Omega_{H}\delta J+ \left(\mathcal{V}-V_{int}\right) \delta P+\Phi_{H} \delta Q.\label{ksc}
\eean
Introducing the mass $\tilde M$ defined by
\bean
\delta \tilde{M}=\delta \mathcal{M}-V_{int}\delta P\label{bhs},
\eean
it follows from \eq{ksc} that
\bean
\delta \tilde{M}=\frac{\kappa}{8\pi}\delta A+\Omega_{H}\delta J+ \left(\mathcal{V}-V_{int}\right) \delta P+\Phi_{H} \delta Q\label{nnm},
\eean
which is the modified first law.
The unknown constants $V_{int}$ will be determined later.
Substituting  \eqs{xbup}, \meq{bne} and \meq{hrs} into \eq{bhs}, we have
\begin{align}
\partial_M \tilde{M}&=\frac{l^4}{\left(-a^2 + l^2\right)^2},\label{xff}\\
\partial_a \tilde{M}&=-\frac{4 a l^4 M}{\left(a^2 - l^2\right)^3},\label{xvh}\\
\partial_l \tilde{M}&=\frac{a^2 l \left(a^2 + 3 l^2\right) M}{\left(a^2 - l^2\right)^3}+\frac{3 V_{int}}{4 l^3 \pi}\label{xel}.
\end{align}
Combining \eqs{xff} and \meq{xvh} and requiring $\tilde{M}=M$ when $a=0$, we have
\bean
\tilde{M} = \frac{M}{\Xi^2}\label{osu}.
\eean
Substituting \eq{osu} into \eq{xel}, we have
\bean
 V_{i n t}=-\frac{4 a^2 l^4 M \pi}{3 \left(a^2 - l^2\right)^2}.
\eean
Define
\bean
V\equiv\mathcal{V}-V_{int}\label{ulu}.
\eean
The first law \meq{nnm} becomes
\bean
 \delta \tilde{M}=\frac{\kappa}{8\pi}\delta A+\Omega_H\delta J+V\delta P+\Phi_{H} \delta Q.\label{pcv}
\eean
Following the same strategy as in subsection \ref{sec3.1}, we obtain the Smarr formula
\begin{equation}
 \tilde{M}=\frac{\kappa A}{4 \pi}+2\Omega J-2P V+\Phi_H Q.\label{eac}
 \end{equation}
Thus, the mass $\frac{M}{\Xi^2}$ satisfies both the first law \meq{nnm} and Smarr formula \meq{eac}.

\subsection{The conformal method}\label{sec4.2}
In this subsection, we shall recover the mass $\frac{M}{\Xi^2}$ by the conformal method.
Similarly to subsection \ref{sec3.2}, we take the conformal factor $\Omega=\frac{1}{r}$. Following almost the same derivation as in subsection \ref{sec3.2}, we have
 \begin{equation}
Q\left[\frac{\partial}{\partial T}\right]=\frac{l}{8 \pi} \oint_{\infty} \mathring{{\mathcal{E}}} _{ab} \left(\frac{\partial}{\partial T}\right)^a \sqrt{\frac{-1}{\mathring{g}^{T T}}}\left[\mathring{g}^{T T}\left(\frac{\partial}{\partial T}\right)^b+\mathring{g}^{\Phi T}\left(\frac{\partial}{\partial\Phi}\right)^b\right] \sqrt{\mathring{g}_{\theta \theta} \mathring{g}_{\Phi\Phi}} d\theta d\Phi.\label{hzj}
\end{equation}
Through \eq{pzx}, we have
\begin{align}
\mathring{{\mathcal{E}}}_{TT}&=-\frac{M \left(a^2 - 2 l^2 + a^2 \cos{2 \theta}\right) \left(a^2 - 4 l^2 +
   3 a^2 \cos{2 \theta}\right)}{4 l^2 \left(a^2 - l^2\right)^2}+O\left(\frac{1}{r}\right)\label{jpr},\\
   \mathring{{\mathcal{E}}}_{T\Phi}&=-\frac{3 \left(a M \left(a^2 - 2 l^2 + a^2\cos{2 \theta}\right) \sin^2\theta\right)}{2\left(a^2-l^2\right)^2}+O\left(\frac{1}{r}\right)\label{lfd}.
\end{align}
 Substituting \eqs{jpr}, \meq{lfd},  \meq{kop} and $\mathring{g}_{\mu \nu}=\left(\frac{1}{r}\right)^2 g_{\mu \nu}$ into \eq{hzj}, we have
  \begin{equation}
Q\left[\frac{\partial}{\partial T}\right]=\frac{M}{\Xi^2}. \label{plm}
\end{equation}
Thus, the conformal mass and the thermodynamic mass are consistent.

\section{Conclusions}\label{sec6}
Since the Kerr-AdS spacetime is not asymptotically flat, there are ambiguities in the mass of black hole. To solve the problem, we  have used the Iyer-Wald formalism to drive the first law \meq{vnpe}
of charged Kerr-AdS black hole  in the extended phase space. Compared to previous treatments, the important improvement is that the variation of the mass in the first law is independent of the choice of the integral surface, avoiding the  divergence of integral at infinity. Based on this formula, we have investigated the mass in different cases. In fact, there are two sources leading to the ambiguities. First, the calculation of mass can be performed in two coordinate systems corresponding to rotating and non-rotating observers in the  spacetime. Second, in each coordinate system, the timelike Killing vector field $\ppn{}{t}$ parallel to the observer is not unique, making the masses different. We have proposed that a physically acceptable mass must satisfy the first law of black hole thermodynamics. By this natural criteria, we have found that for each family of observers, there is a unique timelike Killing vector field, giving rise to an integrable mass in the first law. For rotating observers, the mass is $M/\Xi^{3/2}$ and for non-rotating observers, the mass is $M/\Xi^2$. These results have been confirmed by the conformal method. The advantage of the conformal mass is that the mass can be obtained directly from the integral at the conformal boundary. However, as we have mentioned in the introduction, this method cannot guarantee that the mass satisfies the first law. Only after the appropriate Killing time is chosen by the thermodynamic method  can we obtain the correct mass. Thus, our work clarifies the ambiguities on the mass in Kerr-AdS spacetime.

\section*{Acknowledgements}
This research was supported by NSFC Grants No. 11775022 and 11873044. We thank Prof. Shuang-Qing Wu and Dr. Di Wu for calling our attention to Ref.\cite{Chrusciel:2015sna} and \cite{Wu:2022xpp}. We also thank Prof. Xiao-Kai He for helpful suggestions.

\appendix
\section{Variation of the Lagrangian}\label{appb}
\eqs{bxpo}-\meq{yw} are obtained by varying the Lagrangian \meq{cre}. These are standard calculations which mostly can be found in  Iyer-Wald's original papers  \cite{Iyer:1994ys,Iyer:1995kg} or the textbook \cite{Wald:1984rg}.
For the benefit of the readers, and in order to make the article more self contained, we shall show some intermediate steps as follows.

\eq{cre} can be rewritten as
 \begin{equation}
\mathbf{L}=\frac{1}{16 \pi}\left(\mathbf{L}_{GR}+\mathbf{L}_{E M}+\mathbf{L}_{\Lambda}\right) ,\label{ce1}
\end{equation}
where $\mathbf{L}_{GR}=R \boldsymbol{\epsilon}$, $\mathbf{L}_{E M}=-F^{ab}F_{ab}\boldsymbol{\epsilon}$, $\mathbf{L}_{\Lambda}=-2\Lambda\boldsymbol{\epsilon}$.
The variation of $\mathbf{L}_{GR}$ is given by (see equation E.1.18 in Ref.\cite{Wald:1984rg})
\begin{equation}
\delta \mathbf{L}_{GR}=\delta (R \boldsymbol{\epsilon})=\boldsymbol{\epsilon} \nabla^a v_a+\boldsymbol{\epsilon}\left(R_{a b}-\frac{1}{2} R g_{a b}\right) \delta g^{a b} \text {, }\label{u11}
\end{equation}
where
\begin{equation}
v_a \equiv \nabla^b \delta g_{a b}-g^{b c} \nabla_a \delta g_{b c}=g^{b c}\left(\nabla_c \delta g_{a b}-\nabla_a \delta g_{b c}\right).
\end{equation}

Let $\boldsymbol{e}$ be the fixed volume element associated with a coordinate system and $g_{\mu\nu}$ be the coordinate basis components of the metric. We rewrite
$\mathbf{L}_{E M}$  as
\begin{equation}
\mathbf{L}_{E M}=-\sqrt{-g} \boldsymbol{e} g^{a c} g^{b d} F_{a b} F_{c d} ,\label{lkp}
\end{equation}
where $g$ is the determinant of $(g_{\mu\nu})$. Varying \eq{lkp} leads to
\begin{equation}
\begin{aligned}
\delta \mathbf{L}_{E M} & =(-g^{a c} g^{b d} F_{a b} F_{c d} \delta \sqrt{-g}-\sqrt{-g} g^{b d} F_{a b} F_{c d} \delta g^{a c}-\sqrt{-g} g^{a c} F_{a b} F_{c d} \delta g^{b d} \\
& -\sqrt{-g} g^{a c} g^{b d} F_{a b} \delta F_{c d}-\sqrt{-g} g^{a c} g^{b d} F_{c d} \delta F_{a b})\boldsymbol{e}.
\end{aligned}
\end{equation}
Using $\delta \sqrt{-g}=\frac{\sqrt{-g}}{2} g^{a b} \delta g_{a b}$, we have
\bean
\delta \mathbf{L}_{E M}\eqn F^{c d} F_{c d} \frac{\boldsymbol{\epsilon}}{2} g_{a b} \delta g^{a b}-2 \boldsymbol{\epsilon} F_{a c} F_b{}^c \delta g^{a b}-2 \boldsymbol{\epsilon} g^{a c} g^{b d} F_{c d}  \delta F_{a b}. \\
\eqn -8 \pi \boldsymbol{\epsilon} T_{a b}^{E M} \delta g^{a b}-2 \boldsymbol{\epsilon} F^{a b} \delta F_{a b}. \label{mkls}
\eean
where \eq{klh} has been used in the last step.

Note that
\begin{equation}
F_{a b}=2 \nabla_{[a} A_{b]}, \label{fabt}
\end{equation}
where $A_a$ is the vector potential.
Substitution of \eq{fabt} into \eq{mkls} yields
\begin{equation}
\delta \mathbf{L}_{E M}=-8 \pi \boldsymbol{\epsilon} T_{a b}^{E M} \delta g^{a b}-4 \boldsymbol{\epsilon} F^{a b} \nabla_a \delta A_b .\label{mks}
\end{equation}
Performing the integration by parts on the second term in \eq{mks} gives
\begin{align}
-4 \boldsymbol{\epsilon} F^{a b} \nabla_a \delta A_b  =-4 \boldsymbol{\epsilon} \nabla_a\left(F^{a b} \delta A_b\right)+4 \boldsymbol{\epsilon}\left(\nabla_a F^{a b}\right) \delta A_b.
\end{align}
Therefore
\bean
\delta \mathbf{L}_{E M}=-8 \pi \boldsymbol{\epsilon} T_{a b}^{E M} \delta g^{a b}-4 \boldsymbol{\epsilon} \nabla_a\left(F^{a b} \delta A_b\right)+4 \boldsymbol{\epsilon}\left(\nabla_a F^{a b}\right) \delta A_b .\label{u12}
\eean
Variation of $\mathbf{L}_{\Lambda}$ leads to
\bean
\delta \mathbf{L}_{\Lambda}=-2\delta\left(\Lambda\boldsymbol{\epsilon}\right)=-2\Lambda \delta\boldsymbol{\epsilon}-2\boldsymbol{\epsilon}\delta\Lambda.
\eean
Using $\delta\boldsymbol{\epsilon}=\frac{\boldsymbol{\epsilon}}{2} g^{a b} \delta g_{a b}$, we have
\bean
\delta \mathbf{L}_{\Lambda}=-\Lambda \boldsymbol{\epsilon} g^{a b} \delta g_{a b}-2\boldsymbol{\epsilon}\delta\Lambda .\label{u13}
\eean
Combining \eqs{ce1}, \meq{u11}, \meq{u12} and \meq{u13}, we obtain
\begin{equation}
 \begin{aligned}
\delta\mathbf{L}&=\frac{1}{16 \pi}\boldsymbol{\epsilon}\left(R_{a b}-\frac{1}{2} R g_{a b}-8 \pi T_{a b}^{E M}+\Lambda  g_{a b}\right) \delta g^{a b}+\frac{1}{16 \pi}\boldsymbol{\epsilon} \nabla^a v_a-\frac{1}{4 \pi} \boldsymbol{\epsilon} \nabla_a\left(F^{a b} \delta A_b\right)\\&+\frac{1}{4 \pi} \boldsymbol{\epsilon}\left(\nabla_a F^{a b}\right) \delta A_b -\frac{1}{8 \pi}\boldsymbol{\epsilon}\delta\Lambda .\label{ce12}
\end{aligned}
\end{equation}
Comparing \eq{bxpo} with \eq{ce12}, we find the expressions for $\mathbf{E_g}$ and $\mathbf{E_{A}}$, i.e., \eqs{mtr} and \meq{mtr2}. We also have
\begin{align}
d \boldsymbol{\Theta}&=\frac{1}{16 \pi}\boldsymbol{\epsilon} \nabla^a v_a-\frac{1}{4 \pi} \boldsymbol{\epsilon} \nabla_a\left(F^{a b} \delta A_b\right)\label{kl0}.
\end{align}

The dual form of a 1-form $n_a$ is defined by
\bean
({}^*\boldsymbol{n})_{a b c}=n^d\epsilon_{d a b c}\,.
\eean
Applying the formula
\bean
\boldsymbol{\epsilon}\nabla^a n_a \equiv d {}^*\boldsymbol{n},\label{klaq}
\eean
to \eq{kl0}, we have
\bean
\begin{aligned}
\boldsymbol{\Theta}&=\frac{1}{16 \pi}{}^*\boldsymbol{v}-\frac{1}{4 \pi}{}^*\left(F_{a b} \delta A^b\right)\\&= \frac{1}{16 \pi} \epsilon_{d a b c} g^{d e} g^{f h}\left(\nabla_f \delta g_{e h}-\nabla_e \delta g_{f h}\right)-\frac{1}{4 \pi}\epsilon_{e a b c}F^{ef}\delta A_f,
\end{aligned}
\eean
which is just \eq{yw}.

\section{ Conserved quantities in the conformal method   }\label{appa}
In this appendix, we review the conformal method to calculate conserved quantities\cite{Ashtekar:1999jx}. Consider a $d$-dimensional spacetime denoted by $(\mathbf{M},g_{ab})$ . Suppose there exists a manifold $\mathring{M}$   with boundary $\mathcal{I}$  and $\mathring{M}-\mathcal{I} $ is diffeomorphism to $\mathbf{M}$. $\mathring{M}$  is equipped with a metric $\mathring{g}_{a b}$ and $\mathring{g}_{a b}$ is related to $g_{ab}$ through the conformal transformation $\mathring{g}_{ab}=\Omega^2 g_{ab}$.

Let us start with the Riemann tensor of the metric $\mathring{g}_{ab}$, which can be
decomposed into the Weyl tensor and Ricci tensor
\begin{equation}
\mathring{R}_{a b m n}=\mathring{C}_{a b m n}+\frac{2}{d-2}\left(\mathring{g}_{a[m} \mathring{S}_{n] b}-\mathring{g}_{b[m} \mathring{S}_{n] a}\right),\label{nkl}
\end{equation}
where
\begin{equation}
\mathring{S}_{a b}=\mathring{R}_{a b}-\frac{\mathring{R}}{2(d-1)} \mathring{g}_{a b} .\label{lpa}
\end{equation}
Under the conformal transformation, Ashtekar and  Das obtained
\begin{equation}
S_{a b}=\mathring{S}_{a b}+(d-2) \Omega^{-1} \mathring{\nabla}_a \mathring{n}_b-\frac{1}{2}\left(d-2\right) \Omega^{-2} \mathring{g}_{a b} \mathring{n}^c \mathring{n}_c\label{jkw},
\end{equation}
with $\mathring{n}_a=\mathring{\nabla}_a \Omega$.
Taking the derivative of \eq{jkw} and using \eq{nkl} and Einstein's equation
\begin{equation}
S_{a b}-\frac{\Lambda}{d-1} g_{a b}=8 \pi \left(T_{a b}-\frac{T}{d-1} g_{a b}\right),\label{xkl}
\end{equation}
we have
\begin{equation}
\Omega \mathring{\nabla}_{[a} \mathring{S}_{b] c}+\frac{1}{2}\left(d-2\right) \mathring{C}_{a b c}{ }^d \mathring{n}_d=8 \pi  \tilde{T}_{d[a} \mathring{g}_{b] c} \mathring{n}^d+8 \pi  \mathring{\nabla}_{[a} \left(\Omega \tilde{T}_{b] c}\right),\label{fsx}
\end{equation}
where
\begin{equation}
\tilde{T}_{a b}=T_{a b}-[T /\left(d-1\right)] g_{a b}.
\end{equation}
Note that contracted Bianchi identity on $(\mathring{M}, \mathring{g}_{ab})$ is
\bean
\mathring{\nabla}_{[a} \mathring{R}_{b d] c}{}^d=0.\label{lqs}
\eean
Substituting \eqs{nkl} and \meq{lpa} into \eq{lqs}, we have \cite{Ashtekar:1999jx}
\begin{equation}
\mathring{\nabla}^d \mathring{C}_{a b c d}+\frac{2(d-3)}{(d-2)}  \mathring{\nabla}_{[a} \mathring{S}_{b] c}=0. \label{ndz}
\end{equation}
Eliminating the $\mathring{S}_{ab}$ term in \eq{fsx} by using \eq{ndz} and define
  \begin{equation}
\mathring{K}_{a b c d}:=\lim _{\rightarrow \mathcal{I}} \Omega^{3-d} \mathring{C}_{a b c d},
\end{equation}
we have
\begin{equation}
\mathring{\nabla}^d \mathring{K}_{a b c d} \widehat{=} \lim _{\rightarrow \mathcal{I}}-\frac{2\left(d-3\right)}{\left(d-2\right)} \Omega^{2-d} 8 \pi \left[\tilde{T}_{d[a} \mathring{g}_{b] c} \mathring{n}^d+\mathring{\nabla}_{[a}\left(\Omega \tilde{T}_{b] c}\right)\right],
\end{equation}
where $\widehat{=}$ denotes equality restricted to the boundary $\mathcal{I}$.
By projecting this equation via $\mathring{n}^a \mathring{n}^c \mathring{h}^b{ }_m$ with the induced metric $\mathring{h}_{a b} \widehat{=} \mathring{g}_{a b}-l^2 \mathring{n}_a \mathring{n}_b$ on the boundary and defining $\mathring{\mathcal{E}}_{a b} \widehat{=} l^2 \mathring{K}_{a m b n} \mathring{n}^m \mathring{n}^n$ and $\mathbf{T}_{a b}=\lim_{\rightarrow \mathcal{I}} \Omega^{2-d} T_{a b}$,  Ashtekar and  Das obtained \cite{Ashtekar:1999jx}
\bean
D^d \mathring{\mathcal{E}}_{m d}=-8 \pi (d-3)\mathbf{T}_{a b} \mathring{n}^a \mathring{h}^b{ }_m ,\label{fxc}
\eean
where $D$ is the intrinsic derivative operator on $\mathcal{I}$ compatible with $\mathring{h}_{ ab}$.
Contracting \eq{fxc} with the Killing vector $\xi^m$ on the boundary $\mathcal{I}$ yields
\bean
\xi^m D^d \mathring{\mathcal{E}}_{m d}=-8 \pi(d-3) \mathbf{T}_{a b}\xi^m \mathring{n}^a \mathring{h}^b{ }_m. \label{fvz}
\eean
Using integration by parts, the Killing equation $D_d \xi_m=D_{\left[d\right.} \xi_{\left.m\right]}$ on the boundary, and $\mathring{\mathcal{E}}_{m d}=\mathring{\mathcal{E}}_{\left(m d\right)}$, we find
\bean
\xi^m D^d \mathring{\mathcal{E}}_{m d}=D^d \left(\mathring{\mathcal{E}}_{m d}\xi^m\right).\label{xls}
\eean
Substituting $d=4$, \eq{xls} and $\mathring{h}_{a b} \widehat{=} \mathring{g}_{a b}-l^2 \mathring{n}_a \mathring{n}_b$ into \eq{fvz}, we have
\bean
D^d \left(\mathring{\mathcal{E}}_{m d}\xi^m\right)=-8 \pi \mathbf{T}_{a b}\xi^a \mathring{n}^b,\label{bnf}
\eean
where $\xi^a \mathring{n}_a=0$ has been used.
Integrating \eq{bnf} on the boundary region $\Delta \mathcal{I}$  bounded by two cross sections $C_1$ and $C_2$,  and applying Stokes' theorem, we obtain
\bean
\int_{C_1}\mathring{\mathcal{E}}^{m d}\xi_m \mathring{\epsilon}_{dab}-\int_{C_2}\mathring{\mathcal{E}}^{m d}\xi_m\mathring{\epsilon}_{dab}=\int_{\Delta \mathcal{I}}-8 \pi \mathbf{T}_{a b}\xi^a \mathring{n}^b\,\mathring{\epsilon}_{cde},\label{xla}
\eean
where $\mathring{\epsilon}_{abc}$ is the volume element on the boundary $\mathcal{I}$. Notice the fact that $\mathring{n}^a \mathring{n}_a \widehat{=} \frac{-2 \Lambda}{\left(d-2\right)\left(d-1\right)} = \frac{1}{l^2}$, we should replace $\mathring{n}^b$ with $l\mathring{n}^b$, such that $\mathring{\epsilon}_{bcde}=l\mathring{n}^b\wedge\mathring{\epsilon}_{cde}$. Thus we can rewrite \eq{xla} as
\bean
l\int_{C_1}\mathring{\mathcal{E}}^{m d}\xi_m \mathring{\epsilon}_{dab}-l\int_{C_2}\mathring{\mathcal{E}}^{m d}\xi_m\mathring{\epsilon}_{dab}=\int_{\Delta \mathcal{I}}-8 \pi \mathbf{T}^{a b}\xi_a \mathring{\epsilon}_{bcde}.\label{hxc}
\eean
When $\mathbf{T}_{a b}\xi^a \mathring{n}^b=0$, \eq{hxc} gives the conserved quantity
\bean
Q[\xi]=-\frac{l}{8\pi}\int_{C}\mathring{\mathcal{E}}^{m d}\xi_m \mathring{\epsilon}_{dab},
\eean
which is equivalent to \eq{mk}.
$\xi=\frac{\partial}{\partial t}$ and $\xi=-\frac{\partial}{\partial\phi}$ give the mass and angular momentum, respectively.

\end{document}